
\magnification \magstep1
\raggedbottom
\openup 2\jot
\voffset6truemm
\headline={\ifnum\pageno=1\hfill \else
{\it HAMILTONIAN STRUCTURE OF A FRW UNIVERSE WITH TORSION}
\hfill \fi}
\centerline {\bf HAMILTONIAN STRUCTURE OF A}
\centerline {\bf FRIEDMANN-ROBERTSON-WALKER}
\centerline {\bf UNIVERSE WITH TORSION}
\vskip 1cm
\centerline {GIAMPIERO ESPOSITO}
\vskip 1cm
\noindent
{\it Department of Applied Mathematics and Theoretical
Physics, Silver Street, Cambridge CB3 9EW, U.K.}
\vskip 0.3cm
\noindent
{\it St. John's College, Cambridge CB2 1TP, U.K.}
\vskip 1cm
\noindent
{\bf Summary.} - We study a $R^{2}$ model of gravity with
torsion in a closed Friedmann-Robertson-Walker universe. The
model is cast in Hamiltonian form subtracting from the original
Lagrangian the total time derivative of $f_{K}f_{R}$, where
$f_{K}$ is proportional to the trace of the extrinsic curvature
tensor, and $f_{R}$ is obtained differentiating the Lagrangian
with respect to the highest derivative. Torsion is found to lead
to a primary constraint linear in the momenta and a secondary
constraint quadratic in the momenta, and the full field
equations are finally worked out in detail. Problems to be
studied for further research are the solution of these equations
and the quantization of the model. One could then try to study
a new class of quantum cosmological models with torsion.
\vskip 100cm
\leftline {\bf 1. - Introduction.}
\vskip 1cm
A central issue in modern cosmology is the role played by theories
of gravitation other than Einstein's general relativity in
getting a better understanding of the early universe. Among the
various alternative theories, the ones with torsion [1] are still
receiving careful consideration. Most theories with torsion are
such that spin can be thought of as the source of torsion. There
are at least two very important motivations for studying this
kind of gravitational theories with torsion (we are indebted to
C. Stornaiolo for explaining these issues). In fact:

a) The holonomy theorems [2,3] imply that torsion and curvature
are related, respectively, to the groups of translations and of
homogeneous transformations in the tangent vector spaces to a
manifold. In general relativity the spin does not play any
dynamical role, so that the infinitesimal holonomy group is
given just by the homogeneous transformations. The introduction
of torsion related to spin introduces a much stronger link
between gravitation and particle physics, because it extends
the holonomy group to the translations. A very enlightening
discussion of gauge translations can be found in [4]
(together with the literature given therein) and in [5].
In particular, the introduction of [5] makes clear from the
very beginning the main geometric role played by the
translations in the gauge group: they change a principal fibre
bundle having no special relationship between the points on
the fibres and the base manifold into the bundle of linear
frames of the base manifold. A different view is expressed
in sect. {\bf 5} of [6].

b) At the very high densities present in the early universe,
the effects of spin are no longer negligible. In addition,
theoretical investigations have shown that torsion may lead
to the avoidance of the big-bang singularity in some cases
(see for example [7]).

A good treatment of a theory where the sources of gravity are
mass and spin is for example [8]. In that paper the
Poincar\'e gauge theory of gravitation is studied, setting
up a first-order Lagrangian formalism in a Riemann-Cartan
space-time. However, in our paper we will not focus our
attention on this kind of theories. In fact, we are more
interested in getting a preliminary understanding of some
basic features of the canonical structure of
$R+R^{2}$ theories of gravity with torsion. The reader
should keep in mind that in theories with torsion derivatives
higher than $2$ are not present if frame and connection
are regarded as independent variables. The $R^{2}$ terms
arise already in the torsion-free case as quantum corrections
to the effective action. In general such theories suffer from
serious problems due to the presence of ghosts. However in
[9-11] it has been found that some $R+R^{2}$ theories are
ghost-free. This is indeed a delicate point. In fact in
studying teleparallel theories (where the constraint of
vanishing curvature is imposed), the authors in [12] found
that the most general quadratic Lagrangian for such theories
leads to dipole ghosts. These are field modes with
free-field energies not bounded from below, and exist in view
of $p^{4}$-poles. The authors in [12] also studied the general
Poincar\'e gauge theory, and in sect. {\bf 5} of their paper
they discussed in detail the difference between their work
and the approach used in [9] in dealing with $p^{4}$-poles.
Other reasons for studying $R+R^{2}$ theories with torsion are:

i) The calculation of the Hamiltonian for $R+R^{2}$ models
with torsion is a very interesting field of research, because
it provides an extremely fertile interplay between Dirac's
theory of constrained Hamiltonian systems and the problems of
gravitation and cosmology. Here we would like to mention the
important paper [13], where Dirac's method is applied to the
most general, parity-conserving Lagrangian, which is at most
quadratic in the torsion and the curvature (the special
$R^{2}$ case is treated in [14], whereas higher-order theories
in the Riemannian case are studied for example in [15]).

ii) In the torsion-free case, $R+R^{2}$ theories have been
shown to lead to an inflationary universe [16,17]. Moreover,
other theories with torsion can lead to inflation (see for
example [18,19]).

iii) In $R+R^{2}$ theories in the most general case, torsion
may be expected to propagate, a remarkable property which,
however, is not shared by the Einstein-Cartan-Sciama-Kibble
theory based on spin.

In sect. {\bf 2} we study a $R^{2}$ model of gravity with
torsion in a closed Friedmann-Robertson-Walker (hereafter
referred to as FRW) universe. The consideration of this
model is suggested by a Lagrangian studied in [10]. We
cast the model in Hamiltonian form using a canonical variable
obtained differentiating the Lagrangian with respect to the
highest derivative. Torsion makes itself manifest in that it
induces one primary constraint $\phi_{1}$, linear in the
momenta. Requiring the preservation in time of $\phi_{1}$,
we end up with a secondary constraint $\phi_{2}$ quadratic in
the momenta. Contact is also made with the canonical structure
of the corresponding torsion-free model. In sect. {\bf 3} we
summarize the results obtained, and we mention the problems
to be studied for further research.
\vskip 1cm
\leftline {\bf 2. - $R^{2}$ theories with torsion: a closed
FRW model.}
\vskip 1cm
We are here interested in a $R^{2}$ theory of gravity with
torsion. The Lagrangian of our model is assumed to be given by
$16\pi G L_{g}=N\sqrt{h} \; \mu
{\Bigr({ }^{(4)}R \Bigr)}^{2}$, where no assumption on the sign
of $\mu$ is made. Indeed, a slightly more general Lagrangian
would be
$$
16 {\pi} G L_{g}=N \sqrt{h}
\biggr(\lambda \Bigr({ }^{(4)}R \Bigr)
+\mu {\Bigr({ }^{(4)}R \Bigr)}^{2}\biggr) \; .
$$
This model was at first studied by Rauch and Nieh [10], though
not in a FRW universe and not using canonical techniques. Thus
our Lagrangian is a special case of the Lagrangian studied in
[10] when $\lambda=0$. In so doing our analytical derivations
will be simplified, which in turn implies an advantage in
trying to solve numerically the field equations, which will be
shown to be rather involved. Moreover, setting $\lambda=0$ we
realize a more direct comparison with the canonical structure
of the torsion-free model studied in sect. {\bf 3} of [17].
Thus both our formalism and the physical content of our model
are rather different from what has been discussed in [10]. The
only basic ideas we strictly need to recall are the following.
In a theory of gravity with torsion we have a four-dimensional
space-time manifold with a metric tensor and a nonsymmetric
linear connection
$\Gamma_{\; \; \mu \nu}^{\lambda}$ which obeys the
metricity condition [1]. The torsion tensor in a coordinate
frame is defined by
$$
S_{\mu \nu}^{\; \; \; \lambda}={1\over 2}
\Bigr(\Gamma_{\; \; \mu \nu}^{\lambda}
-\Gamma_{\; \; \nu \mu}^{\lambda}\Bigr)
=-S_{\nu \mu}^{\; \; \; \lambda} \; .
\eqno (2.1)
$$
One also defines a contorsion tensor
$$
C_{\nu \mu}^{\; \; \; \lambda}=S_{\nu \mu}^{\; \; \; \lambda}
-S_{\mu \; \; \nu}^{\; \; \lambda}
+S_{\; \; \nu \mu}^{\lambda} \; ,
\eqno (2.2)
$$
so that
$$
\Gamma_{\; \; \mu \nu}^{\lambda}
=\left\{{\lambda \atop \mu \nu}\right\}
-C_{\nu \mu}^{\; \; \; \lambda} \; ,
\eqno (2.3)
$$
where $\left\{{\lambda \atop \mu \nu}\right\}$
are the Christoffel symbols. We shall use a convention
according to which the curvature tensor is given by [20]
$$
R_{\; \; \mu \nu \beta}^{\alpha}=
\Gamma_{\; \; \mu \nu , \beta}^{\alpha}
-\Gamma_{\; \; \mu \beta , \nu}^{\alpha}
+\Gamma_{\; \; \mu \nu}^{\lambda} \;
\Gamma_{\; \; \lambda \beta}^{\alpha}
-\Gamma_{\; \; \mu \beta}^{\lambda} \;
\Gamma_{\; \; \lambda \nu}^{\alpha} \; .
\eqno (2.4)
$$
We will focus our attention on a closed FRW universe. In view
of the hypothesis of spatial homogeneity and isotropy, we are
interested in a torsion tensor $S_{\mu \nu}^{\; \; \; \lambda}$
whose only nonvanishing components are
$$
S_{10}^{\; \; \; 1}=S_{20}^{\; \; \; 2}
=S_{30}^{\; \; \; 3}=Q(t) \; .
\eqno (2.5)
$$
In our model, the metric may be locally cast in the form
$$
ds^{2}=-N^{2}(t)dt^{2}+a^{2}(t) \Bigr(d\chi^{2}
+(\sin \chi)^{2}d\theta^{2}
+(\sin \chi)^{2}(\sin \theta)^{2}d\phi^{2}\Bigr) \; .
\eqno (2.6)
$$
Thus one finds (see appendix)
$$
{ }^{(4)}R={6\over a^{2}}
+12{\left({{\dot a}\over Na}\right)}^{2}
-{72{\dot a}Q\over aN^{2}}+{96 Q^{2}\over N^{2}}
+{6\over N}{d\over dt}
\left({{\dot a}\over Na}-{2Q\over N}\right) \; .
\eqno (2.7)
$$
We now define
$$
I={\mu \over 16 {\pi} G}
\int_{M}{\Bigr({ }^{(4)}R\Bigr)}^{2}N \sqrt{h} \; d^{4}x \; .
\eqno (2.8)
$$
Therefore, putting $\beta=3{\pi}/2G$, $\alpha={\rm lg}(a)$,
${\widetilde \mu}=\mu \beta$, we find
$$
I=\int L \; dt \; ,
\eqno (2.9)
$$
$$ \eqalignno{
L&={{\widetilde \mu}\over 12}\exp[3\alpha]N \cdot \cr
& \cdot {\left[6\exp[-2\alpha]
+{12 {\dot \alpha}^{2}\over N^{2}}
-72{{\dot \alpha}Q\over N^{2}}
+{96 Q^{2}\over N^{2}}
+{6\over N}{d\over dt}
\left({{{\dot \alpha}-2Q}\over N}\right)\right]}^{2}
\; .
&(2.10)\cr}
$$
Thus, putting
$$
{Q\over N}=y \; , \; \; \; \;
\tau \int N \; dt \; ,
\eqno (2.11)
$$
denoting by a prime the derivative with respect to $\tau$, and
defining $\mu^{0}=3{\widetilde \mu}$, we obtain the relations
$$
I=\int {\widetilde L} \; d\tau \; ,
\eqno (2.12)
$$
$$
{\widetilde L}=\mu^{0}\exp[3\alpha]
{\Bigr[\exp[-2\alpha]+2{\alpha '}^{2}-12y \alpha'
+16y^{2}+{\alpha}''-2y' \Bigr]}^{2} \; .
\eqno (2.13)
$$
Now, in the torsion-free case (compare with [17]) one defines the
variable $z \equiv \partial {\widetilde L}/ \partial \alpha''$,
and one considers the Lagrangian (we are indebted to D. Giulini
for clarifying this point):
$$
L' \equiv {\widetilde L}-{d\over d\tau}({\alpha '} z) \; .
\eqno (2.14)
$$
The geometrical meaning of $z$ is that it is proportional to
the scalar curvature through ${\widetilde \mu} \; \exp[3\alpha]$.
In defining (2.14), we fix $\alpha$ and $z$ rather than
$\alpha$ and $\alpha'$ at the initial and final times. In our
model with torsion, we can no longer use (2.14) as our final
Lagrangian $L'$. In fact, if one defines
$p_{\alpha}=\partial L' / \partial \alpha'$,
$p_{y}=\partial L' / \partial y'$ and
$p_{z}=\partial L' / \partial z'$ using (2.14), the resulting
relations do not involve $y'$, whereas $p_{y}=-2z$. Thus
it is not possible to compute the Legendre transform:
$H=p_{\alpha}\alpha'+p_{z}z'+p_{y}y'-L'$. However, we can
point out that in (2.14) $\alpha'$ is a function proportional
to the trace of the extrinsic curvature tensor. In our model
with torsion, $\alpha'$ gets replaced by $\alpha'-2y$
(see (2.7) and (2.13)). This suggests to define
$$
L' \equiv {\widetilde L}-{d\over d\tau}
\Bigr[(\alpha' -2y)z \Bigr] \; .
\eqno (2.15)
$$
In deriving (2.15) in a more systematic way, we can point out that
the total derivative appearing in (2.14) is
$$
\alpha'' {\partial {\widetilde L}\over \partial \alpha''}
+ \alpha' {d\over d\tau}{\partial {\widetilde L}\over
\partial \alpha''} \; .
$$
Thus, if torsion has to play a role, we expect having to add the term
$$
y'{\partial {\widetilde L}\over \partial y'}
+y{d\over d\tau}{\partial {\widetilde L}\over \partial y'}
=-2{d\over d\tau}(yz) \; ,
$$
so that we end up with definition (2.15). To sum up, in defining
the correct $L'$ we require that

a) it must be possible to compute the Legendre transform:
$H=p_{\alpha}\alpha'+p_{z}z'+p_{y}y'-L'$;

b) $L'$ must reduce to (2.14) in the torsion-free case, when
$y=0$;

c) $L'$ must not contain derivatives of $y$ higher than $y'$,
because these derivatives are absent in (2.13);

d) $L'$ must be defined in a unique way, through a
general method;

e) $L'$ must be of the kind $L' \equiv {\widetilde L}
-(d/d\tau)(f_{K}f_{R})$ where $f_{K}$ is the function
$\alpha'-2y$, proportional to the trace $K$ of the full
extrinsic curvature tensor, and $f_{R}$ is proportional
through ${\widetilde \mu} \; \exp[3\alpha]$ to the full
scalar curvature ${ }^{(4)}R$.

It is also worth remarking that:

i) in view of (2.1)-(2.3) and (2.5), the only components of
the connection (2.3) which depend both on $\alpha$ and on
$Q$ are $\Gamma_{01}^{1}$, $\Gamma_{02}^{2}$ and
$\Gamma_{03}^{3}$, and they are all equal to
${\dot \alpha}-2Q=N(\alpha'-2y)$;

ii) formula (2.15) can be written in the form
$$
L' \equiv {\widetilde L}-{d\over d\tau}
\left(x' {\partial {\widetilde L}\over \partial x''} \right)
$$
which is formally of the same kind of (2.14), if we define
$x=\alpha-2\int y \; d\tau$;

iii) if we require that the constraint $p_{y} \approx 0$ should
be avoided and that all constraints produced by torsion should
be linear in the momenta, we are led to define
$$
L' \equiv {\widetilde L}-{d\over d\tau}
\left[\alpha' \biggr(z+\sum_{l=1}^{\infty}f_{l}(\alpha)
y^{l} \biggr) \right] \; .
$$
But in so doing there are infinitely many forms of $L'$, one
for each choice of the set of functions $\{f_{l}(\alpha)\}$,
and our construction is completely arbitrary, violating
requirement d).

Thus we will use (2.15), where, putting
$$
y=u' \; ,
\eqno (2.16)
$$
one has
$$
z \equiv {\partial {\widetilde L}\over \partial x''}
=2\mu^{0} \exp[3\alpha]
\biggr[\exp[-2\alpha]+2{\alpha'}^{2}-12u'\alpha'
+16{u'}^{2}+\alpha''-2u'' \biggr] \; .
\eqno (2.17)
$$
As one can easily check, putting $y=u'$ one has the advantage
of dealing with less involved calculations. Using the definitions
$p_{\alpha}=\partial L' / \partial \alpha'$,
$p_{z}=\partial L' / \partial z'$,
$p_{u}=\partial L' / \partial u'$, we find that the effective
Hamiltonian ${\widetilde H}$ is given by
$$
{\widetilde H}=H+\gamma \phi_{1} \; .
\eqno (2.18)
$$
In (2.18), one has
$$ \eqalignno{
H&=p_{\alpha}\alpha'+p_{z}z'+p_{u}u'-L' \cr
&=-4zp_{z}^{2}+{p_{z}p_{u}\over 2}
+{z^{2}\exp[-3\alpha]\over 4\mu^{0}}
-z\exp[-2\alpha]
&(2.19)\cr}
$$
with the primary constraint
$$
\phi_{1}=p_{u}+2p_{\alpha}-4zp_{z} \; .
\eqno (2.20)
$$
The constraint $\phi_{1} \approx 0$ must now be preserved in
time using Dirac's method [21] for constrained Hamiltonian
systems. Namely, computing the Poisson bracket of $\phi_{1}$
with ${\widetilde H}$, we find the secondary constraint:
$$
\phi_{2}=16zp_{z}^{2}-2p_{u}p_{z}+{7z^{2}\over 2\mu^{0}}
\exp[-3\alpha]-8z\exp[-2\alpha] \; .
\eqno (2.21)
$$
In other words, torsion introduces a primary constraint
$\phi_{1}$ linear in the momenta, which in turn leads to one more
secondary constraint $\phi_{2}$ quadratic in the momenta.
Finally, so as to preserve in time the constraint
$\phi_{2} \approx 0$, we require that the following Poisson
bracket must vanish:
$$
\left\{\phi_{2},{\widetilde H}\right\}
=\left\{\phi_{2},H\right\}
+\gamma \left\{\phi_{2},\phi_{1}\right\} \; .
\eqno (2.22)
$$
Setting the right-hand side of (2.22) equal to zero and
solving for $\gamma$ we find
$$
\gamma=-{\left\{\phi_{2},H \right\} \over
\left\{\phi_{2},\phi_{1}\right\}} \; ,
\eqno (2.23)
$$
where
$$
\left\{\phi_{2},\phi_{1}\right\}
=64zp_{z}^{2}-8p_{u}p_{z}-{49 z^{2}\over \mu^{0}}
\exp[-3\alpha]+64z \exp[-2\alpha] \; ,
\eqno (2.24)
$$
$$ \eqalignno{
\left\{\phi_{2},H\right\}&=
\left(96 \exp[-2\alpha]z-{72\exp[-3\alpha]\over \mu^{0}}
z^{2}\right)p_{z} \cr
&+\left({9\exp[-3\alpha]\over 2\mu^{0}}z
-6\exp[-2\alpha]\right)p_{u} \; .
&(2.25)\cr}
$$
Using the constraint $\phi_{1} \approx 0$, we can also
cast (2.19) in the form
$$
H=-2zp_{z}^{2}-p_{\alpha}p_{z}
+{z^{2}\exp[-3\alpha]\over 4\mu^{0}}
-z\exp[-2\alpha] \; .
\eqno (2.26)
$$
Thus the equations of motion of our model are given by
$$
\alpha'={\partial {\widetilde H}\over \partial p_{\alpha}}
=-p_{z}+2\gamma \; ,
\eqno (2.27)
$$
$$
z'={\partial {\widetilde H}\over \partial p_{z}}
=-4zp_{z}-p_{\alpha}-4\gamma z \; ,
\eqno (2.28)
$$
$$
u'={\partial {\widetilde H}\over \partial p_{u}}
=\gamma \; ,
\eqno (2.29)
$$
$$
p_{\alpha}'=-{\partial {\widetilde H}\over \partial \alpha}
={3z^{2}\over 4\mu^{0}}\exp[-3\alpha]
-2z \exp[-2\alpha] \; ,
\eqno (2.30)
$$
$$
p_{z}'=-{\partial {\widetilde H}\over \partial z}
=2p_{z}^{2}-{z\exp[-3\alpha]\over 2\mu^{0}}
+\exp[-2\alpha]+4\gamma p_{z} \; ,
\eqno (2.31)
$$
$$
p_{u}'=-{\partial {\widetilde H}\over \partial u}
=0 \; ,
\eqno (2.32)
$$
plus the constraints: $\phi_{1} \approx 0$,
$\phi_{2} \approx 0$ and the formula for $\gamma$
(see (2.19)-(2.21) and (2.23)-(2.25)). In other words, we
deal with a coupled system of six first-order ordinary
differential equations subject to constraints. A remarkable
result is that we have written the field equations in
Hamiltonian form without having to solve any differential
equation involving torsion. Moreover, (2.32) implies that
$p_{u}={\rm const}$. It is also very important remarking
that, imposing the {\it particular value} $H=0$ (see (2.19)),
we can solve for $zp_{z}^{2}$. Inserting this relation into
the constraint $\phi_{2}=0$, we obtain the relation
$$
\exp[\alpha]-{3\over 8 \mu^{0}}z=0 \; .
\eqno (2.33)
$$
In other words, requiring the constraints $\phi_{1}=0$,
$\phi_{2}=0$ and setting $H=0$ is equivalent to require
that $\Phi_{1}$, $\Phi_{2}$ and $H$ must vanish,
where $\Phi_{1}=\phi_{1}$ and $\Phi_{2}$ is given by the
left-hand side of (2.33). The constraints $\phi_{1}$ and
$\phi_{2}$ are second class (they {\it cannot} be brought into
the first class by suitable linear combinations [21]).
It is also worth emphasizing that in the torsion-free case
no primary constraints such as $\phi_{1}$ are found to arise.
Setting $y=0$ in (2.13) and using (2.14) one finds then
$$
\alpha'=-p_{z} \; ,
\eqno (2.34)
$$
$$
z'=-4zp_{z}-p_{\alpha} \; ,
\eqno (2.35)
$$
$$
p_{\alpha}'={3z^{2}\over 4\mu^{0}}\exp[-3\alpha]
-2z\exp[-2\alpha] \; ,
\eqno (2.36)
$$
$$
p_{z}'=2p_{z}^{2}-{z\over 2\mu^{0}}\exp[-3\alpha]
+\exp[-2\alpha] \; ,
\eqno (2.37)
$$
plus the particular value for the Hamiltonian here chosen
$$
2zp_{z}^{2}+p_{\alpha}p_{z}+z\exp[-2\alpha]
-{z^{2}\over 4\mu^{0}}\exp[-3\alpha]=0 \; .
\eqno (2.38)
$$
So as to avoid confusion, it is important to remark that,
using our notation, one has
$$
\alpha'={{\dot a}\over Na} \; , \; \; \; \;
\alpha''={1\over N}{d\over dt}
\left({{\dot a}\over Na}\right) \; ,
$$
whereas in other papers such as [17] one defines
$$
\alpha'={d\alpha \over d\eta}
={d\alpha \over dt}{dt \over d\eta}
=a {\dot \alpha}={\dot a} \; ,
$$
$$
\alpha''={d\alpha' \over d\eta}=a{d\alpha' \over dt}
=a {\ddot a} \; .
$$
System (2.27)-(2.32) with the constraints (2.20)-(2.21) is
very difficult, and at the classical level we have not a
criterion which picks out a preferred choice of initial
conditions. Indeed, exact solutions for theories with torsion
are known for a large class of theories. An important recent
paper is [22], where the authors have studied the Hamiltonian
structure of Poincar\'e gauge theory. In the case of spherical
symmetry and for the charged Taub-NUT metric, the authors
of [22] have obtained the most general torsion configuration
for a large class of quadratic Lagrangians (see also references
in [22] for related work). However, our original aim was just
the study of the canonical structure of the model described by
(2.5)-(2.6) and (2.8) at the classical level. Thus we prefer
to stop here our analysis, and to outline in detail the
unsolved issues in the next section.
\vskip 1cm
\leftline {\bf 3. - Conclusions.}
\vskip 1cm
In this paper we have studied a Hamiltonian approach to a closed
FRW universe with torsion. The consideration of our $R^{2}$
Lagrangian (see (2.8)) has been suggested by a model at first
studied by Rauch and Nieh [10]. Our main results are the
following:

1) the technique used by Horowitz [17] in the torsion-free
case can be generalized using (2.15)-(2.17);

2) torsion modifies the canonical structure of the theory in
that it induces a primary constraint $\phi_{1}$ linear in the
momenta and a secondary constraint $\phi_{2}$ quadratic in the
momenta (see (2.20)-(2.21));

3) the field equations are much more complicated, and are given
by (2.27)-(2.32).

Thus the most remarkable feature of our model seems to be the
appearance of the additional constraints $\phi_{1}$ and
$\phi_{2}$, or equivalently $\Phi_{1}=\phi_{1}$
and $\Phi_{2}$ (see (2.33)). In general relativity, the
secondary constraints reflect the invariance of the theory
under four-dimensional diffeomorphisms. Thus in our model
we can say that torsion introduces an additional invariance under
four-diffeomorphisms, leading to $\phi_{2}$, which contributes
to the generation of the dynamics through $\gamma$
(see (2.23)-(2.25) and (2.27)-(2.32)). In the as yet
undeveloped minisuperspace model, we expect the physical state
$\psi$ will still be a function of the three-metric $h_{ij}$
and of the extrinsic curvature $K_{ij}$ as in the torsion-free
case [17]. In fact, the nonvanishing part of torsion is in our
case just the one contained in $K_{(ij)}$ (see appendix of [20]),
and from (2.27) and (2.29) one has $\alpha'-2u'=
\alpha'-2y=-p_{z}$, which is formally identical to (2.34),
because in our model with torsion the trace $K$ is proportional
to $\alpha'-2y$ (see sect. {\bf 2}). Indeed, Hamiltonian
methods had also been used for the canonical quantization of
more complicated $R^{2}$ theories of gravity in the
torsion-free case [23,24], and in studying spherically
symmetric gravitational fields with electromagnetism and a
massless scalar field as sources [25]. For theories with torsion,
a rather important paper is [26], and in fact the ansatz (2.5)
of our paper is closely related to the ansatz studied in sect.
{\bf 2} and sect. {\bf 6} of [26]. A minor difference between
our work and [26] lies in the different parametrization of the
FRW metric, and in the fact the authors of [26] set the lapse
function equal to one, whereas we keep it arbitrary. But the
main difference is the following: the authors of [26] do not
perform a canonical analysis, and thus they do not deal with
six first-order equations subject to constraints. A comparison
with their analytical derivation has not yet been possible
because it is very difficult to solve our equations of
motion. Therefore, the main problems to be studied for
further research are the choice of the initial conditions for
the numerical integration of the field equations, the
quantization of the model using path integrals and the attempt
of applying the Hartle-Hawking proposal [27] or other
proposals. In fact we cannot solve the field equations without
knowing the initial conditions. Quantum cosmology can provide
a way for doing so, but then one has to work out the
path-integral quantization, discussing in detail the measure in
the path integral, the metrics we are summing over, the gauge
fixing and the ghost action (namely the action involving
Faddeev-Popov ghosts). Moreover, it is necessary to
understand whether the underlying quantum theory is affected by
negative-norm states, whose occurrence can be a serious drawback.
The quantization procedure is here complicated by the additional
constraints $\Phi_{1}$ and $\Phi_{2}$.

Indeed, torsion is much less {\it ad hoc} than the fundamental
scalar fields considered so far in cosmological models (for
the role played by scalar fields, see for example [28,29]),
though its existence is still to be proved. Moreover, in the
torsion-free case $R+R^{2}$ theories may play an important role
not only for the inflationary era, or in giving a self-consistent
probabilistic interpretation of the wave function of the
universe [30], but also in completely different processes such
as black-hole evaporation [31]. This is why we think it is
important to try to solve the above-mentioned problems. To be
honest, we also must say that it is not yet clear to which extent
we can use a closed FRW metric so as to study the early universe.
Therefore the model we have studied is just a mathematical
idealization. Nevertheless, the derivations obtained show that
it is worth studying theories of gravity with torsion in further
detail from a Hamiltonian point of view. We hope that our paper
can stimulate further interaction between canonical gravity,
theories with torsion and quantum cosmology. We think it would be
also very useful to study the singularity problem for $R^{2}$
theories with torsion both in the isotropic and in the
anisotropic case.
\vskip 1cm
\centerline {$* \; * \; *$}
\vskip 1cm
We are grateful to P. D'Eath, D. Giulini and J. Louko for
useful conversations and to St. John's College for financial
support.
\vskip 1cm
\leftline {\it APPENDIX}
\vskip 1cm
We are interested in the formula for the scalar curvature
${ }^{(4)}R$ for a theory of gravity with torsion. This kind of
calculation is performed in detail in [32]. However, this
reference is not easily available. Thus we prefer to rely
on the appendix of [20]. In that appendix, Pilati works out
the Lagrangian density ignoring a total divergence. Now,
in general form, and bearing in mind eqs. (A.16)-(A.18) of
[20], we can write
$$ \eqalignno{
{ }^{(4)}R&=K_{(ab)}K^{(ab)}-K_{[ab]}K^{[ab]}+K^{2}
-2 \tau_{(ab)}K^{(ab)}+2\tau_{[ab]}K^{[ab]}
+2K\tau-2{{\dot K}\over N}\cr
&+{ }^{(3)}R+A+B+C \; ,
&(A.1)\cr}
$$
where
$$ \eqalignno{
A&=h^{ij}\left[\biggr((C_{mi}^{\; \; \; m})_{,j}
-(C_{ji}^{\; \; \; m})_{,m}
+C_{ji}^{\; \; \; l}C_{ml}^{\; \; \; m}
-C_{mi}^{\; \; \; l}C_{jl}^{\; \; \; m}\biggr) \right. \cr
&\left. +{ }^{(3)}\left\{{l \atop im}\right\}
C_{jl}^{\; \; \; m}
+{ }^{(3)}\left\{{m \atop lj}\right\}
C_{mi}^{\; \; \; l}
-{ }^{(3)}\left\{{l \atop ij}\right\}
C_{ml}^{\; \; \; m}
-{ }^{(3)}\left\{{m \atop lm}\right\}
C_{ji}^{\; \; \; l}\right] \; ,
&(A.2)\cr}
$$
$$
B=-{2\over N}\Bigr(h^{ij}N_{\mid ij}
-N^{i}K_{,i}\Bigr) \; ,
\eqno (A.3)
$$
$$
C=-\rho^{a}q_{a}+2\rho_{\; \; \mid a}^{a}
-2\rho^{a}\rho_{a} \; .
\eqno (A.4)
$$
In (A.4), $\rho^{a}$ and $q_{a}$ are the ones defined in
(A.30)-(A.31) of [20]. Using (2.5)-(2.6) and (A.1)-(A.4)
of our appendix, one finds (2.7). Other useful references
for the Hamiltonian formulation of ECSK theory are [33-35].
Finally, we wish to remark that, putting
$$
S={2h^{ki}S_{k0i}\over N} \; ,
$$
(A.1) leads to a surface term involving torsion in the
action integral
$$
{1\over 8{\pi}G} \int_{\partial M}S\sqrt{h} \; d^{3}x
$$
in addition to the usual one of Einstein's theory [36-38].
\vskip 10cm
\leftline {\it REFERENCES}
\vskip 1cm
\item {[1]}
F. W. HEHL, P. von der HEYDE, G. D. KERLICK and J. M. NESTER:
{\it Rev. Mod. Phys.}, {\bf 48}, 393 (1976).
\item {[2]}
A. TRAUTMAN: {\it Symp. Math.}, {\bf 12}, 139 (1973).
\item {[3]}
C. STORNAIOLO: Ph. D. Thesis (University of Naples, 1987).
\item {[4]}
E. A. LORD and P. GOSWAMI: {\it J. Math. Phys. (N.Y.)},
{\bf 29}, 258 (1988).
\item {[5]}
P. K. SMRZ: {\it J. Math. Phys. (N.Y.)}, {\bf 28}, 2824 (1987).
\item {[6]}
D. IVANENKO and G. SARDANASHVILY: {\it Pramana}, {\bf 29},
21 (1987).
\item {[7]}
M. DEMIANSKI, R. DE RITIS, G. PLATANIA, P. SCUDELLARO
and C. STORNAIOLO: {\it Phys. Lett. A}, {\bf 116}, 13 (1986).
\item {[8]}
J. D. Mc CREA: in {\it Differential Geometric Methods in
Mathematical Physics}, in {\it Proceedings of the XIV International
Conference}, edited by P. L. GARCIA
and A. P\'EREZ-REND\'ON
(Springer-Verlag, Berlin, 1987), p. 222.
\item {[9]}
E. SEZGIN and P. van NIEUWENHUIZEN: {\it Phys. Rev. D},
{\bf 21}, 3269 (1980).
\item {[10]}
R. RAUCH and H. T. NIEH: {\it Phys. Rev. D}, {\bf 24},
2029 (1981).
\item {[11]}
M. GASPERINI: {\it Phys. Lett. B}, {\bf 205}, 517 (1988).
\item {[12]}
R. KUHFUSS and J. NITSCH: {\it Gen. Rel. Grav.},
{\bf 18}, 1207 (1986).
\item {[13]}
I. A. NIKOLIC: {\it Phys. Rev. D}, {\bf 30}, 2508 (1984).
\item {[14]}
V. SZCZYRBA: {\it Phys. Rev. D}, {\bf 36}, 351 (1987).
\item {[15]}
V. SZCZYRBA: {\it J. Math. Phys. (N.Y.)}, {\bf 28},
146 (1987).
\item {[16]}
S. W. HAWKING and J. C. LUTTRELL: {\it Nucl. Phys. B},
{\bf 247}, 250 (1984).
\item {[17]}
G. T. HOROWITZ: {\it Phys. Rev. D}, {\bf 31}, 1169 (1985).
\item {[18]}
R. DE RITIS, P. SCUDELLARO and C. STORNAIOLO:
{\it Phys. Lett. A}, {\bf 126}, 389 (1988).
\item {[19]}
A. J. FENNELLY, J. C. BRADAS and L. L. SMALLEY:
{\it Phys. Lett. A}, {\bf 129}, 195 (1988).
\item {[20]}
M. PILATI: {\it Nucl. Phys. B}, {\bf 132}, 138 (1978).
\item {[21]}
P. A. M. DIRAC: {\it Lectures on Quantum Mechanics}
(Belfer Graduate School of Science, Yeshiva University,
New York, N.Y., 1964).
\item {[22]}
P. BAEKLER and E. W. MIELKE: {\it Fortschr. Phys.},
{\bf 36}, 549 (1988).
\item {[23]}
D. BOULWARE in: {\it Quantum Theory of Gravity}, edited by
S. M. CHRISTENSEN (Adam Hilger, Bristol, 1984), p. 267.
\item {[24]}
I. L. BUCHBINDER and S. L. LYAHOVICH:
{\it Class. Quantum Grav.}, {\bf 4}, 1487 (1987).
\item {[25]}
B. K. BERGER, D. M. CHITRE, V. E. MONCRIEF and Y. NUTKIN:
{\it Phys. Rev. D}, {\bf 5}, 2467 (1972).
\item {[26]}
H. GOENNER and F. MULLER-HOISSEN: {\it Class. Quantum Grav.},
{\bf 1}, 651 (1984).
\item {[27]}
J. B. HARTLE and S. W. HAWKING: {\it Phys. Rev. D},
{\bf 28}, 2960 (1983).
\item {[28]}
S. W. HAWKING: {\it Nucl. Phys. B}, {\bf 239}, 257 (1984).
\item {[29]}
G. ESPOSITO and G. PLATANIA: {\it Class. Quantum Grav.},
{\bf 5}, 937 (1988).
\item {[30]}
M. POLLOCK: {\it Nucl. Phys. B}, {\bf 306}, 931 (1988).
\item {[31]}
R. C. MYERS and J. Z. SIMON: {\it Phys. Rev. D},
{\bf 38}, 2434 (1988).
\item {[32]}
O. ALVAREZ: Senior Thesis (Princeton University, Princeton,
N.J., 1974).
\item {[33]}
L. CASTELLANI, P. van NIEUWENHUIZEN and M. PILATI:
{\it Phys. Rev. D}, {\bf 26}, 352 (1982).
\item {[34]}
R. DI STEFANO and R. T. RAUCH: {\it Phys. Rev. D},
{\bf 26}, 1242 (1982).
\item {[35]}
J. ISENBERG and J. NESTER: in {\it General Relativity
and Gravitation}, edited by A. HELD (Plenum Press, New York,
N.Y., 1980), p. 23.
\item {[36]}
J. W. YORK: {\it Phys. Rev. Lett.}, {\bf 28}, 1082 (1972).
\item {[37]}
G. W. GIBBONS and S. W. HAWKING: {\it Phys. Rev. D},
{\bf 15}, 2752 (1977).
\item {[38]}
J. W. YORK: {\it Found. Phys.}, {\bf 16}, 249 (1986).

\bye